# On Security Strategies for Addressing Potential Vulnerabilities in 6G Technologies Deployable in Healthcare

Chinazunwa Uwaoma
*Center for Information Systems and Techhnology*
*Claremont Graduate University*
California, USA
chinazunwa.uwaoma@cgu.edu

*Abstract*— Researchers are now focusing on 6G as a new network technology that will bring significant gains over the previous generations while many sectors are still implementing the 5G network in their business processes and operations. Meanwhile, key technological fields that will be influenced by 6G networks have been identified. These include distributed artificial intelligence, intelligent radio, real-time intelligent edge computing, and 3D intercoms. Additionally, each area and potential application of 6G is supported by relevant emerging technologies. Nevertheless, these 6G technology and applications have significant security vulnerabilities that must be addressed before the complete adoption of 6G networks. The healthcare is one of the sectors that are benefiting from the great features introduced in the 5G networks that enhance digital communications and data protection; that notwithstanding, there are still security flaws in 5G technologies that can be transferred to the 6G networks if not properly addressed. This paper highlights the key areas of 6G networks that would provide grand support for the development of healthcare systems. It also identifies certain vulnerabilities in the previous cellular networks that are transferable to 6G networks, and suggests security strategies including zero trust initiatives that could be implemented to address the security concerns.

## I. Introduction

There is no doubt the healthcare system is becoming one of the prime targets of cybersecurity threats and attacks in recent times. The reason being that digital healthcare system involves the use of mobile phone and other connected devices for care delivery, and these devices are susceptible to cyber-attacks given their resource constraints. The evolution of cellular network technologies however, has continued to improve digital communications in healthcare system, the latest being the 5G network with great Internet of Things - complaint features like Ultra-Reliable Low Latency Communications (URLLC) that allows for the simultaneous delivery of high-quality services to an increasing number of connected devices [1].

The Internet of Things (IoT), Big Data (BD), supercomputing, artificial intelligence (AI) techniques like machine learning (ML) and deep learning, as well digital security tools like blockchain, have all contributed to the extraordinary opportunity to build an integrated ecosystem for new opportunities in the healthcare and other industries. These advancements are capable of addressing some of the most pressing issues facing policymakers and providers of healthcare services, such as the provision of universal, equitable, and sustainable healthcare. Further, integration of these technologies has the potential to fundamentally alter disease screening, diagnosis, and monitoring, as well as to improve and personalize treatments by allowing for more precise disease progression profiling [2].

A new era of network technology ushers in new and distinctive solutions and applications. Despite the fact that some applications from earlier network generations will still be used in 6G networks, the 6G era appears to hold a lot of promises in the applications areas of multi-sensory extended reality (XR) and wireless brain-computer interactions (BCI) for digital healthcare and wellbeing [1].

Healthcare delivery is changing as a result of the use of big data analytics and cloud computing, as well as the shift to consumer-centric healthcare. The IoT and the use of wearable medical devices, also referred to as the Internet of Medical Things (IoMT), are two concepts that are associated with the use of cloud computing. With the massive amounts of data these innovations have produced, the promise of better patient care, better clinical data, increased efficiency, and lower costs has also been realized. However, many of these modern connected devices lack proper management and security. These limitations can impact the actual care of patients in addition to patients' data that are vulnerable to cybersecurity threats and attacks [2].

The post-COVID pandemic surge in telemedicine and telehealth services gave rise to a new category of patient services, admissions, and treatments. As cloud computing and other emerging technologies are increasingly being adopted in the healthcare industry, the risks of cybersecurity attacks will reflect frequency rates that are comparable to, if not higher than, those observed in the industry's current dominant legacy technology platforms [2].

Many security strategies have been proposed to address the evolving cyberthreat landscape in the digital estate. One of such strategies is the zero trust (ZT) framework which has been widely adopted in both private and public networks, and it is considered a formidable security approach to handle most of the security vulnerabilities associated with previous cellular network generations. The ZT model is also a promising strategy of modernizing IoT security without limiting the scope of the IoT systems.

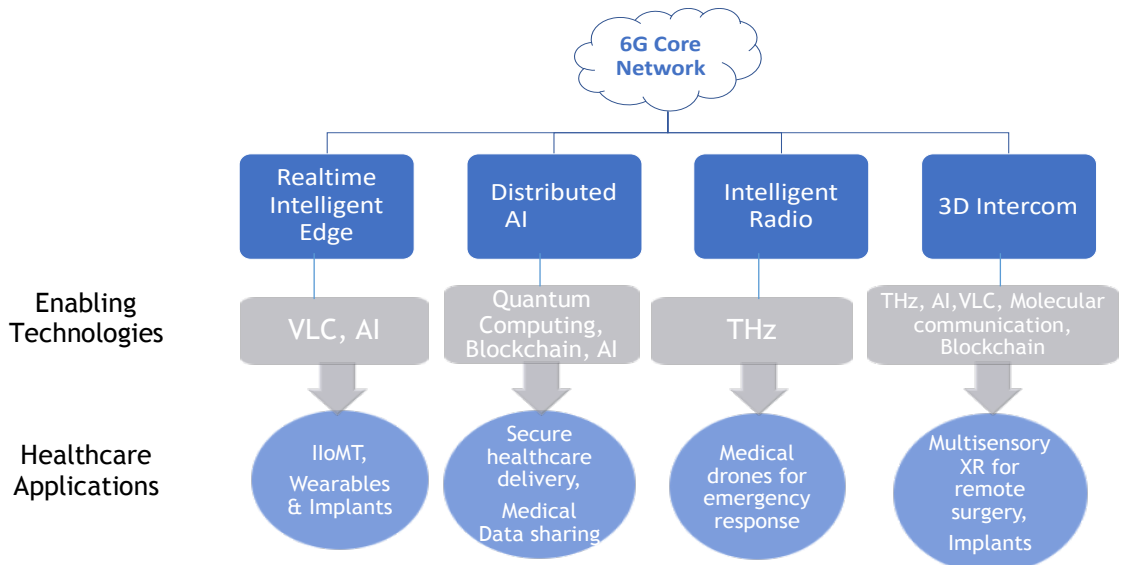

Fig. 1. Key Areas of 6G Networks and Technologies that support Healthcare Systems

Healthcare breaches are more likely because of the current cybersecurity gaps in healthcare and the lack of qualified IT workers in the health industry [2]. The adoption of 6G networks and the related technologies may offer healthcare providers special opportunities to modernize their care and delivery systems and integrate security and privacy capabilities that would have been impossible with earlier network technologies, if done effectively. The key contributions of this paper are:

- Highlighting and examining key areas of 6G technologies that apply to healthcare systems.

- Identifying potential vulnerabilities in 6G technologies and applications that can impact healthcare systems, as well as recommendation of security measures that could be adopted to protect healthcare systems from being exploited by threat actors.

- Presenting an expanded ZT framework with integrated 6G technological features to enhance healthcare security.

The rest of the paper is structured as follows: Section II highlights the key areas of 6G networks that are applicable to healthcare systems. Section III reviews 6G enabling technologies and healthcare applications. Potential vulnerabilities in 6G technologies that might impact healthcare systems are discussed in Section IV alongside the recommended security measures. Section V concludes the discussions.

## II. ASPECTS OF 6G NETWORKS APPLICABLE TO HEALTHCARE SYSTEMS

It is anticipated that the 6G networks will support application areas like connected robotics, autonomous systems, multi-sensory XR, wireless BCIs, as well as smart blockchain and distributed ledger applications [1]. With the emerging and enabling technologies like molecular communications, quantum communications, visible light communication (VLC), AI and blockchain can improve healthcare service delivery and communications. For instance, as an improvement on the 5G networks, 6G technologies will support the Internet of Everything (IoE). In other words, they (6G networks) are required to be IoE-compliant as opposed to 5G networks, which are IoT-compliant. This implies that the 6G network will be a significant decentralized system with capability of making smart decisions at various levels [3]. Also, the ability to access and manage a portion of the collected data should be available to every edge device. Additionally, the edge device should be capable of independently computing and storing data. These are all hallmarks of the 6G features that are applicable to wearable medical devices often located at the point of care - network edge. Fig. 1 shows the major aspects of the 6G network, the enabling technologies and possible healthcare systems that can be supported by these technologies.

### A. Real-time Intelligent Edge

The evolution of the 5G wireless communications network opened up numerous possibilities for the use of cutting-edge technologies to meet healthcare needs. Consequently, this has led to the development of new technologies for therapy and telesurgery which has now furthered the growth of telehealth. It has been demonstrated that 5G has incredibly low latency in comparison to the previous network generation, which enables it to transmit at higher frequency. This provides the opportunity to bring all supporting technologies close to the point of care in real time [4]. However, given that the base stations must be densely populated, there is also the issue of signal degradation during communication among connected devices [5].

The 6G network on the on the other hand, brings with it, innovative technologies that do not only have extremely low network latency and real-time intelligence, but are also capable of enabling self-awareness and self-adaptation in network elements and devices. Such interactive AI-powered services and functions are supported in 6G [1]. In addition, applications of new communications like VLC and molecular communication can address signal degradation in communication among interconnected medical devices.

*B. Distributed AI*

AI is considered an important component and serves as the backbone for other aspects of the future network as it provides the capability for making intelligent decision at various granular levels [3]. Although some elements of AI applications have been implemented in 5G networks, the shortcomings of the current framework that was available in the early stages of its development limit the support for AI-driven technologies in 5G [1]. However, the design of 6G network allows the integration of new AI tools and networking functions that provides medical devices and applications located at the edge the ability to not only access and control part of the data, but also to compute and store data independently. When this approach is implemented properly, it is seen a step towards strengthening and improving security and privacy by keeping medical data at edge [3].

Other key areas that will advance the implementation of 6G technologies in healthcare systems are intelligent radio and ubiquitous coverage or 3D intercom network. While an AI-based intelligent radio is expected to handle emergency conditions in real-time that are not provisioned in 5G networks due to latency issues, the 3D intercom is to "provide connectivity in the global scale by enabling wireless coverage to meet the high-altitude and deep-sea communication scenarios that are lacking in current network architecture" based on the existing terrestrial cellular infrastructure of the 5G network [3]. The 3D intercom that provides service at the ground level, space and undersea levels has been broadly discussed in [1], [3], [6]. This essentially presents opportunity for people to access medical and healthcare services in real-time irrespective of the strata of the sphere where they are located. It is expected that 6G will expand the scope and capabilities of 5G with an improvement and introduction of new technologies. Table I shows transferable features of these components in 5G technology that can be improved in the future 6G network.

III. SUPPORTING TECHNOLOGIES FOR 6G APPLICATIONS IN HEALTHCARE SYSTEMS

It is envisioned that the 6G networks will be built on a number of different technologies. These include both current and emerging technologies, like Terahertz (THz) technology, VLC, post-quantum cryptography, molecular communication, and enhanced edge computing technologies, distributed ledgers (DL) specifically, the blockchain. Healthcare applications that will benefit from these technologies include Intelligent Internet of Medical Things (IIoMT) for remote surgery, autonomous medical drones, connected ambulance, and implants.

*A. TECHNOLOGIES*

*1) Artificial Intelligence:* Although artificial intelligence in the 5G network is essentially controlled in remote locations with access to enormous volumes of training data and strong but private computing centers, AI will become a more integral part of the 6G network [1]. Currently, some IoT sensors and implants are unable to meet the necessary computational requirements due to resource constraints. Development of intelligent 6G-enabled IoT networks however, will need device-based AI training methods on nano IoT devices and integrated wearable technology for services like intelligently enhanced life assistance. Additionally, mobile and intelligent healthcare devices need to be able to carry out AI tasks for real-time edge computing and data exchange [7].

*2) Blockchain:* There are numerous applications of Blockchain technology in a 6G network. For instance, decentralized networks, DL technologies, and spectrum sharing could enhance network performance and make network management easier [1]. These blockchain features can also be applied to medical data sharing among hospitals and healthcare delivery organizations (HDOs) to ensure anonymity and integrity of sensitive data.

*3) Quantum Communication:* Quantum computing technology has significant potentials for deployment in 6G networks [1]. With the right technological advancements, quantum communication could theoretically offer complete security and should be ideal for long-distance communication. As part of the new 6G-enabled IoT networks, quantum communication can support traffic data up to 1 Gb $s^1/m^2$ to meet the high data rate requirements and massive device density in a given location, which can provide smart healthcare applications with extremely high reliability [7].

*4) Terahertz Technology:* Despite being extensively used in 5G networks, mm-wave bands are insufficient for 6G due to the need for high transmission rates [1]. In comparison to the mm-wave band, the 0.1–10 THz band is used for terahertz communication because it has more spectrum resources available for transmission. Utilizing the THz band has a number of advantages, one of which is the ability of THz communication technology to support data rates of 100 Gbps or more [3]. This new technology would enable 6G-IoT networked ambulances that can be used by doctors and other medical professionals to provide medical assistance and prompt intervention in order to save lives in emergency situations [7].

*5) Visible Light Communication:* A promising solution to the rising demand for wireless connectivity is visible light communication technology. As a complementary technology to Radio Frequency (RF), it is possible that this area will see advancements in wireless communication, microelectronics, and energy harvesting. For example, IoT devices for smart healthcare can use this technology to provide low-power and low-cost communication. Additionally, sensors that are implanted in the body can harvest solar energy from outside light to monitor medical conditions [7].

*6) Molecular Communication:* In living things with nanoscale structures, molecular communication is a common natural occurrence [1]. A molecular communication signal is created and transmitted with very little energy consumption. The technology for molecular communication has great potentials for 6G communications, even though it is still in its early stage. Utilizing biochemical signals to transmit data is the main principle of molecular communication. By identifying various pathogens and biomarkers related to deadly and communicable diseases without the need for external power, new materials like "MXenes-based intelligent biosensors can further revolutionize the implants" [7].

TABLE I. TRANSFERABLE FEAUTURES IN 5G TECHNOLOGIES AND THEIR IMPROVEMENTS IN 6G

| Key Components | Transferable Features in 5G | Improvement in 6G |
|---|---|---|
| Realtime Intelligence Edge | *Mobility, End-to-End Reliability, Applications, and Data Rate* Support for Mobility (Up to 500 km h$^{-1}$), Improved mobile broadband, enabling URLLC), Peak data rate (20Gbps), and end-to-end latency (1 − 5 ms) [7], [8] | Support for Mobility (Up to 1000 km h$^{-1}$) Combines URLLC and Enhanced mobile broadband, (eMBB), enabling Mobile Broad Bandwidth and Low Latency (MBBLL) Support for massive machine type communication (mMTC) Peak data rate (1 Tbps), and end-to-end latency (<1 ms) [7], [8] |
| Distributed AI | *Connected Devices* Centralized IoT-compliant, Improved connectivity and coverage for smart devices | Distributed IoE-compliant, Very high device connectivity density |
| Intelligent Radio | *Spectral Efficiency and Bandwidth* Channel Modeling and Spectrum sharing, Spectral Efficiency of about 30 bps Hz$^{-1}$ [8] | Spectral Efficiency of about 100 bps Hz$^{-1}$ [8] |
| 3D Intercom | N/A | Localization Precision, Support for aerial vehicle and Autonomous Drone systems |

## B. APPLICATIONS

*1) Multi-sensory Extended Reality (XR) Applications:* The 5G user experience for virtual reality (VR) and augmented reality (AR) has already been improved by the 5G networks' high bandwidth and low latency [1]. Healthcare applications using 6G will enhance VR/AR experiences through the use of multiple sensors to gather sensory information and give users feedback. There will be less restriction on surgical procedures due to time and distance. It is envisioned that IIoMT will use tactile and holographic communication as well as augmented and virtual reality to aid physicians to remotely perform complicated procedures or surgery [7].

*2) Connected Robotics and Autonomous Drone Systems:* It is anticipated that 6G networks will make it possible to create and deploy connected robotics that integrate AI logic into the network's structure as well as embedded intelligence across the entire network, reducing the need for human intervention in industrial processes. With the help of robotic devices with millisecond latency and extremely high reliability, 6G robotics can be used to implement remote surgery, thus, allowing doctors to perform it from a distance [7]. Due to the limitations of 5G networks, an autonomous drone system has yet to be fully deployed. However, 6G networks may be able to realize the full potential of such systems. In the event of an emergency, Unmanned Aerial Vehicles (UAVs) can reduce the time it takes for medical personnel to arrive by delivering medical supplies (such as surgical instruments, medications, etc.) from hospitals located in various places. This prevents delays caused by road traffic and allows for prompt assistance.

*3) Wireless Brain-Computer Interactions:* The main idea of wireless brain-computer interactions is to create a link between the brain and a machine. Wireless BCI has primarily been used to help disabled people control assistive technology, but a new approach to BCI has been developed to accelerate spelling proficiency using brain signals. BCI work in a manner akin to XR applications, but they are more reliant on the physical world and require an assurance of "Quality-of-Physical-Experience" [1].

## IV. POTENTIAL VULNERABILITIES IN 6G TECHNOLOGIES CRITICAL TO HEALTHCARE SYSTEMS

Every cellular network generation has its own security shortcomings. The challenge of updating fundamental protocols increases vulnerability despite the existence of numerous measures to reduce exploitation [9]. Poor authentication and resource limitations are two issues that affect all cellular network generations and are challenging to solve. Denial of service (DoS) attacks on authentication servers, distributed denial of service (DDoS) attacks on signaling, energy depletion attacks, and user tracking are also examples of potential attacks on 6G security architecture and applications [10].

### A. Progression of Security and Privacy Issues in Cellular Networks

The threat landscape in wireless systems has expanded due to more sophisticated attacks and skilled attackers. The 1G network was designed to deliver voice services and uses analog signals. However it lacks a specified wireless standard [1]. The security and privacy issues in this network include unencrypted telephone services, unauthorized access as well as eavesdropping and cloning attacks [11]. Network services supported in 2G network include enabling voice and short messaging services, providing anonymity of users' identity as well as privacy solution and radio path encryption [9]. However, the network is flawed by unauthorized access and one-way authentication issues, end-to-end encryption problem, as well as traceability and eavesdropping attacks. The 3G network was developed in the year 2000 to provide internet access and advanced services that were not provisioned in 1G and 2G. It also supports various privacy considerations including secure location and identification. However, there are security and privacy issues identified in this network which include vulnerability and threats related to Internet Protocol (IP) traffic, poor encryption techniques, integrity threats, unauthorized data and service access, and DoS attacks [9].

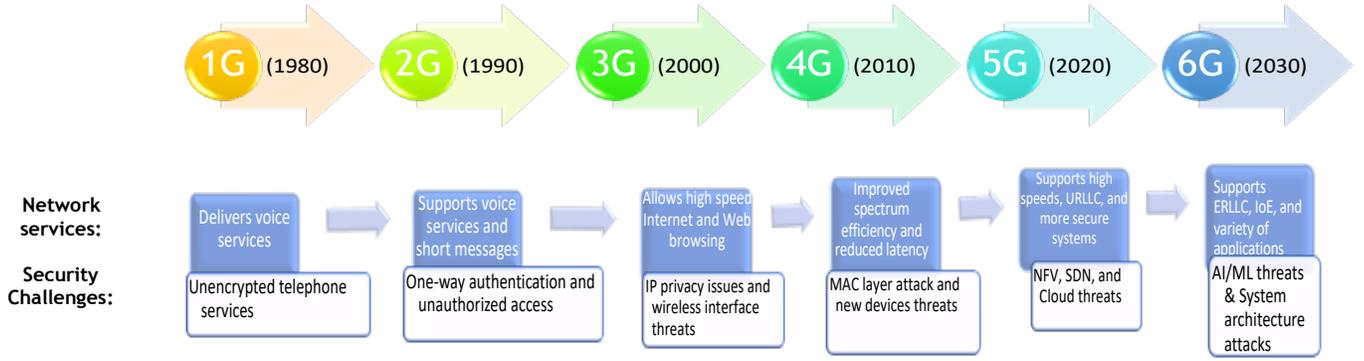

Fig. 2. Security and Privacy Issues in Each Generation of Cellular Networks

Remarkable improvements in the 4G network over the 3G include handling sophisticated applications and supporting a variety of intelligent mobile devices. This network generation also offers increased network transmission and communication speeds. However, issues with network entity authentication, eavesdropping, data insertion or deletion, wireless link security including Medium Access Control (MAC) layer vulnerabilities, are the major security flaws in the 4G networks [1], [9]. Enabling network features of the 5G include the capacity to deliver higher-quality services to all network entities while connecting an increasing number of growing devices, support for URLLC, as well as enabling Software-Defined Networking (SDN) and its variant Software-Defined Perimeter (SDP) as well as Network Functions Virtualization (NFV) [12], [13]. However, security concerns that have been identified in this network generation are network resource attacks, and massive MIMOs used to mask both passive and active spying. Also, implementing OpenFlow as part of SDN makes users more vulnerable to malicious applications or activities [14]. Fig. 2 highlights security issues in the development of the previous cellular networks as well as potential security shortcomings in the 6G network.

### B. Security Measures to Protect Against Possible Threats in 6G Technologies Deployable in Healthcare Systems

Potential security and privacy issues that have been identified to affect healthcare applications supported by 6G technologies can be categorized as Malicious behavior (MB), Unauthorized access control (UAC), Data communication (DC), and Data encryption (DE) attacks [1], [8]. This section takes a look at the critical problems and challenges related to security, privacy and trust in 6G network technologies, and also suggests new security approaches that can be adopted to guarantee protection, reliability, and privacy in healthcare systems. Table II summarizes potential vulnerabilities in 6G network that might impact healthcare applications, as well recommended security measures to address the vulnerabilities.

TABLE II. VULNERABILITIES IN 6G THAT COULD IMPACT HEALTHCARE APPLICATIONS AND RECOMMENDED SECURITY MEASURES

| Healthcare Applications | Identified Vulnerabilities and Security Measures | | |
|---|---|---|---|
| | *6G Enabling Technology* | *Potential Security & Privacy Issues* | *Recommended Security Measures* |
| Intelligent Internet of Medical Things (IIoMT), Wearables & Implants | VLC, AI | *UAC, MB, and DC* DoS attacks, hijacking, spoofing attacks, and eavesdropping attacks [7]. | Application of fine-tuned control processes using ML [15] Automatic detection of network anomalies and alerts [16] Implementing decentralized and distributed IoT/IoE networks |
| Secure healthcare delivery, Medical data aggregation and sharing using blockchain. & distributed ledger | Quantum Computing, Blockchain, AI | *DE, DC, and UAC* DDoS attacks, hijacking, spoofing attacks. Data ownership and ethical concerns. | Using ML and post-quantum cryptography techniques [17] Implementing decentralized and distributed AI and Blockchain technologies, Applying intelligent algorithms (e.g. Federated Learning Algorithms) to safeguard user privacy and confidentiality [7] |
| Medical drones for Emergency Response | THz, AI | *MB and AUC* WiFi-based attacks, eavesdropping, hijacking, spoofing, and DoS attacks. | Deployment of unsupervised ML methods to ensure proper authentication processes [18] Adopting Privacy-preserving framework to manage permission issues in UAV networks [1] |
| Multi-sensory XR for remote surgery, Connected ambulance for people with disabilities | THz, AI, VLC, Molecular communication, Blockchain | *AUC, MB, DC, and DE* Possible leak of critical and confidential information. | Using VLC technology to provision secure protocol in communication processes [19] |
| Wireless brain-computer interactions & Implants | AI, VLC, Molecular communication | *MB and DE* Use of hacking applications to access highly-sensitive neurological information [1]. | Implementing a new coding scheme to improve security of data transmission using molecular communication technologies [20] Adopting a password method that requires the user to enter a certain psychological state to prevent reply attacks [1] |

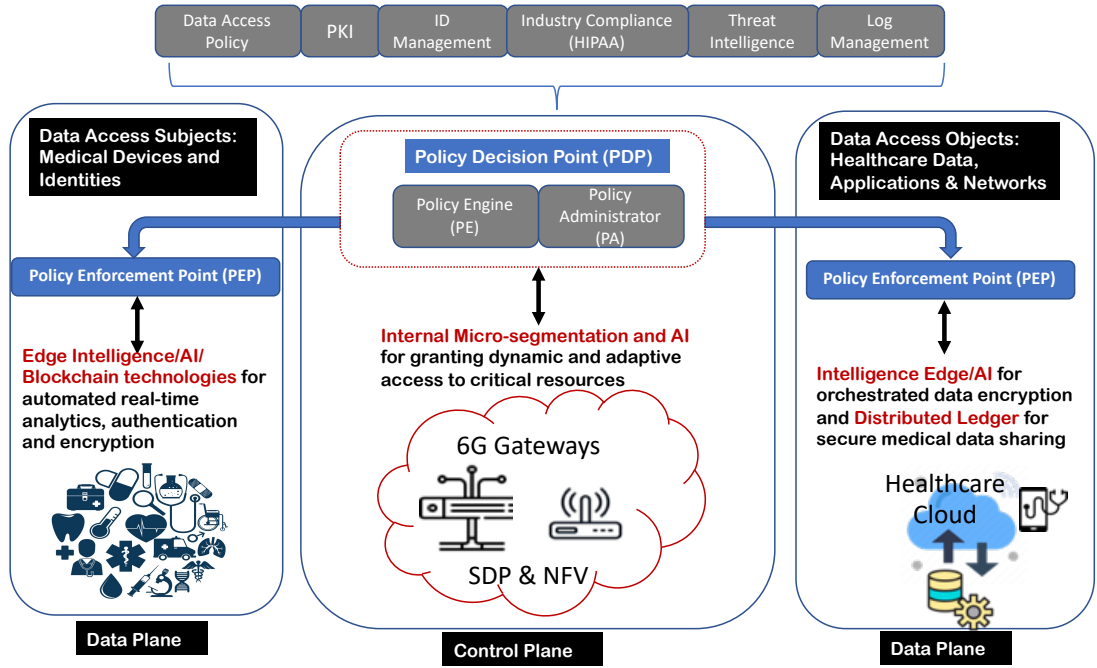

Fig. 3. An expanded Zero Trust Framework that incorporates 6G technological features to enhance healthcare security

## C. Zero Trust Initiatives in Healthcare Services

Additionally, in order to address security concerns related to perimeter-based networks, the 6G security architecture is anticipated to support the fundamental security tenets of zero trust (ZT) in the mobile communication network. ZT is a security approach that prioritizes safeguarding system and network resources above all else. The network architecture is presumed to be accessible from outside or to be unreliable, and it is assumed that an attacker could reside on the network. Network entity relationships, protocol operations, and access rules are all parts of the zero trust architecture (ZTA), a security framework that makes use of the zero-trust concept. It has been suggested that the design of the 6G network architecture incorporates security measures and requirements from ZTA. These include: Virtualized Security Solutions, Automated Management System, Data security using AI, Users' Privacy-preserving, and Post-Quantum Cryptography [9].

Without a doubt, the ZT strategy is receiving more attention in healthcare than ever before because it deals with the difficulties associated with implementing and adopting new technology at a faster rate. However, ZT is considered independent of any infrastructure as it is agnostic to location, application, data, device or technology [21]. ZT implementation in healthcare affects multiple domains such as (a) **Devices** – clinical tools, tablets/phones, IoMT (b) **Identities** – clinical teams, administrative teams, patients. (c) **Applications** – EMRs/EHRs, billings, scheduling (d) **Data** – patient records, lab results, patient details [22].

ZT revolves around identity and access management (IAM), cloud security gateways, as well as applications and network security. It is a common knowledge that medical IoT devices are used to obtain, monitor, and report vital patient information. However, providing unique identity, encryption, and up-to-date inventory of the vast number of these devices can be a daunting task and requires some form of automation to ensure secure healthcare networks. Also, the mobile nature of healthcare workers constitutes another challenge to establish MFA and granular authorization to these set of users [23].

While ZT implementation seems difficult to kickstart, there are unique considerations and simplified processes for any health organization that wants to embark on this journey as outlined. in [23]. All the same, the benefits of ZT outweighs the challenges when properly implemented. The essence of having "total absence of trust" is to limit the attack surface even as boundaries continue to grow increasingly complex [24].

## D. Integrating 6G Technological Features in ZT to enhance Security of Healthcare Systems

Intelligent medical IoT devices used to deliver health care services provide improved patient outcomes, alerts and constant monitoring to reduce labor shortage in the industry. However, they also extend attack surface exploitable by cybercriminals. Further, these devices ship with no built-in security and they are too expensive to replace. Hence, it is critical that device manufacturers include universal support for security certificates that validate and authenticate device-to-network transactions. [25]. There is also the need to incorporate AI, ML, and distributed ledger (blockchain) technologies in ZT to help inform, augment, and strengthen access policies to reduce the odds of a breach. Medical facilities often need to grant third-party access to devices for services and in doing so, outsiders are given permission that exceed the minimum required for their work. ZT framework can enforce the Principle of Least Privilege (PoLP) to ensure that transactions involve only the necessary services, data, and systems thereby limiting access to specific device being serviced [25]. Fig. 3 shows an expansion of the ZT framework to include the afore-mentioned technologies that can improve security in healthcare services.

## V. CONCLUSIONS

The importance of enhancing technology security prior to deployment cannot be overstated because newer network standards perform better in new applications. However, since the current 5G network architecture still has flaws and weaknesses discovered in earlier network generations, long-term design change is required to correct them. This paper highlights the key areas of 6G networks that would provide grand support for the development of healthcare systems. It also identifies certain vulnerabilities in the previous cellular networks that are transferable to 6G networks, and suggests security strategies that could be implemented to address the security concerns. The network concepts and security issues discussed here provides a baseline for future research on developing robust network frameworks and models to safeguard healthcare systems in the era of 6G networks. Though not covered in this paper, potential health hazards resulting from the use of 6G technologies like THz also need to be examined.